\documentclass[graybox]{svmult}

\usepackage{mathptmx}       
\usepackage{helvet}         
\usepackage{courier}        
\usepackage{type1cm}        
\usepackage{makeidx}         
\usepackage{graphicx}        
\usepackage{multicol}        
\usepackage[bottom]{footmisc}
\usepackage{cite}

\usepackage{amsfonts}
\usepackage{amsmath}
\usepackage{sidecap}
\usepackage[affil-it]{authblk}
\usepackage{color}

\makeindex             

\def\kmin{k_{\mathrm{min}}}
\def\pn{p_n}
\def\pr{p_r}
\def\kvec{\mathbf{k}}
\def\Pk{P({\kvec})}
\def\Fkm{F_{\kvec, m}}
\def\skm{s_{\kvec, m}}
\def\skmo{s_{\kvec, m-1}}
\def\dskm{\dot{s}_{\kvec, m}}
\def\sumk{\sum_{\kvec}}
\def\summ{\sum_m}
\def\bs{\beta_s}
\def\Bkm{B_{k, m}}
\def\Bkom{B_{k-1, m}}

\def\ft{f_t}

\newcommand{\be}{\begin{equation}}
\newcommand{\ee}{\end{equation}}

\begin{document}

\title*{Service adoption spreading in online social networks}
\author{Gerardo I\~niguez, Zhongyuan Ruan, Kimmo Kaski, J\'anos Kert\'esz and M\'arton Karsai}
\authorrunning{Gerardo I\~niguez {\it et al.}}
\institute{Gerardo I\~niguez \at Institute for Research in Applied Mathematics and Systems, National Autonomous University of Mexico, 01000 M{\'e}xico D.F., Mexico; Department of Computer Science, Aalto University School of Science, 00076 Aalto, Finland \email{gerardo.iniguez@aalto.fi}
\and Zhongyuan Ruan \at Center for Network Science, Central European University, 1051 Budapest, Hungary, \email{zyruan86@gmail.com}
\and Kimmo Kaski \at Department of Computer Science, Aalto University School of Science, 00076 Aalto, Finland, \email{kimmo.kaski@aalto.fi}
\and J\'anos Kert\'esz \at Center for Network Science, Central European University, 1051 Budapest, Hungary, \email{KerteszJ@ceu.edu}
\and M\'arton Karsai \at Univ de Lyon, ENS de Lyon, INRIA, CNRS, UMR 5668, IXXI, 69364 Lyon, France, \email{marton.karsai@ens-lyon.fr}}

%
%
\maketitle

\abstract{The collective behaviour of people adopting an innovation, product or online service is commonly interpreted as a spreading phenomenon throughout the fabric of society. This process is arguably driven by social influence, social learning and by external effects like media. Observations of such processes date back to the seminal studies by Rogers and Bass, and their mathematical modelling has taken two directions: One paradigm, called simple contagion, identifies adoption spreading with an epidemic process. The other one, named complex contagion, is concerned with behavioural thresholds and successfully explains the emergence of large cascades of adoption resulting in a rapid spreading often seen in empirical data. The observation of real world adoption processes has become easier lately due to the availability of large digital social network and behavioural datasets. This has allowed simultaneous study of network structures and dynamics of online service adoption, shedding light on the mechanisms and external effects that influence the temporal evolution of behavioural or innovation adoption. These advancements have induced the development of more realistic models of social spreading phenomena, which in turn have provided remarkably good predictions of various empirical adoption processes. In this chapter we review recent data-driven studies addressing real-world service adoption processes. Our studies provide the first detailed empirical evidence of a heterogeneous threshold distribution in adoption. We also describe the modelling of such phenomena with formal methods and data-driven simulations. Our objective is to understand the effects of identified social mechanisms on service adoption spreading, and to provide potential new directions and open questions for future research.}

\bigskip

\section{Introduction}
\label{sec:intro}

A human society abounds with examples of collective patterns of behaviour that arise due to the correlated decisions of a large number of individuals. This is evidenced in the spread of religious beliefs and political movements, in the behavioural, cultural, and opinion shifts in a population, in the adoption of technological and medical innovations, in the rise of popularity of political and media figures, in the growth of bubbles in financial markets, and in the use of products and online services. All of these phenomena tend to evolve similarly over time, as they start with individuals that independently from their peers and due to external influence such as mass media, take the risk by adopting a certain behaviour~\cite{valente-thresholds-1996,Toole2012Modeling}. Then, these processes continue as friends, colleagues, and acquaintances observe such individuals and engage with the same behaviour, therefore participating in a spreading process throughout society~\cite{Kleinberg2007Cascading}.

The way ideas, products, and behaviour spread throughout a population over time, commonly know as {\it innovation diffusion}, was first observed empirically in the mid 20th century by the likes of Rogers~\cite{Rogers2003Diffusion} and Bass~\cite{Bass1969}. In the following decades, many mathematical models were introduced with the goal of identifying mechanisms by which behaviour diffuses through society~\cite{Granovetter1978Threshold,schelling1969models,Axelrod1997Dissemination}. One of the first (and arguably simplest) is the Bass model for forecasting sales of new consumer durables~\cite{Bass1969}, which characterises the diffusion of innovation as a process of contagion initiated by some external influence~\cite{Toole2012Modeling} (e.g. mass communication, news media) and promoted by internal, social influence~\cite{Centola2010Spread} (via word-of-mouth, viral marketing, etc.). The model assumes a homogeneous population of adopters and it predicts that aggregated sales data has an s-shaped pattern as a function of time~\cite{Granovetter1978Threshold,Zhang2016Empirically}.

Despite the success of the Bass model and similar diffusion-like models to capture qualitatively the temporal behaviour of adoption processes, macroscopic models only provide empirical generalisations based on the behaviour of society as a whole (by means of aggregated data on adoption rates, for example). Hence, these models do not take into account individual heterogeneities and the complex structure and dynamics of social processes~\cite{Kiesling2012Agent}. In other words, since the same macro-level behaviour may arise from several individual-level mechanisms (like learning, externalities, or contagion), it is difficult for these models to assess what mechanisms are actually responsible for large-scale spreading phenomena~\cite{Goel2012Structure,BorgeHolthoefer2013Cascading,Goel2015Structural}. In order to overcome this issue, agent-based diffusion models consider behavioural heterogeneities, networked social interactions~\cite{RevModPhys.81.591,Bakshy2012The} and decision-making processes based on the cognitive capacities of individuals~\cite{Holt06,bikhchandani-hirshleifer-welch-92, Karsai2014Complex}. Then, behaviour at the level of society emerges dynamically from the interplay between network structure and the actions of people. This microscopic approach allows for the modelling of varying behaviour across individuals, while recognising that social interactions and interpersonal communication are essential in determining adoption~\cite{valente-thresholds-1996,Romero2011Differences}.

Under the network approach, the Bass model is an archetypal example of {\it simple contagion}~\cite{Barrat2008Dynamical} where, akin to the transmission of a disease, information and individuals' willingness to adopt may propagate with exposure to a single person engaging in some particular behaviour. However, when adoption turns out costly, risky or controversial, the spread of ideas and products often requires social reinforcement and exposure to several sources, a phenomenon usually called {\it complex contagion}~\cite{Centola2007Complex,porter2016}. The requirement of multiple interactions for adoption was first implemented theoretically by Granovetter via behavioural thresholds, namely {\it `the number or proportion of others who must make one decision before a given actor does so'}~\cite{Granovetter1978Threshold}. Following this idea various agent-based network models have been introduced and analysed by Watts and others~\cite{valente-thresholds-1996,Watts2002Simple,Handjani1997Survival,Neill2005Cascade,Watts2007Influentials,Melnik2013Multistage,Gomez2010Modeling, Karampourniotis2015The,Miller2015Complex} in order to understand the properties of threshold-driven social contagion.

Despite the allure of social influence as the reason behind innovation diffusion, it is more challenging to identify causal mechanisms in adoption spreading than in biological contagion, since the same empirical, large-scale observations may be obtained as effects of social influence~\cite{Onnela2010Spontaneous}, homophily~\cite{McPherson2001}, or the environment. For example, collective adoption patterns may appear as a consequence of homophilic structural correlations, where interacting individuals adopt due to their similar interests and not due to actual social influence~\cite{porter2016}. Hence distinguishing between the effects of social influence and homophily at the individual level remains a challenge~\cite{Aral2009,Shalizi2011}. Moreover, regarding the particular role social influence may have in adoption spreading, several assumptions have been proposed about its functional dependency on the number of adopters necessary to influence an individual. While Granovetter and others~\cite{Granovetter1978Threshold,Watts2002Simple} suggest a simple linear dependency, as observed in some large techno-social systems \cite{Karsai2014Complex}, Latan\'e~\cite{Latane1981The} argues for non-linear effects that have been demonstrated empirically by online experiments at different scales~\cite{Centola2010Spread,Centola2011An,Suri2011Cooperation}.

Perhaps one of the most intriguing features of threshold-driven social contagion is its ability to capture what Watts calls the {\it robust yet fragile} nature of complex systems~\cite{Watts2002Simple}. This means that a population may be robust and disregard many ideas and products, but suddenly exhibit fast system-wide adoption patterns known as {\it behavioural cascades}. While homophily suggests that adoption behaviour is only seemingly correlated, and simple contagion implies that external influence always induces global adoption in a connected population, complex contagion captures the additional feature that large cascades of behavioural patterns tend to happen only rarely, and may be triggered by actions at the individual level that are indistinguishable from the rest. Indeed, behavioural cascades are rare but potentially disrupting social spreading phenomena, where collective patterns of exposure arise through reinforcement as a consequence of small initial perturbations~\cite{Motter2017Unfolding}. Examples include the rapid emergence of political and grass-root movements~\cite{GonzalezBailon2011Dynamics,BorgeHolthoefer2011Structural,EllisInformation}, or the fast spreading of information \cite{Goel2012Structure,Watts2007Influentials,Dow2013Anatomy,Gruhl2004Information,Banos2013Role,Hale2013Regime,Leskovec2005Patterns,Leskovec2007Dynamics} and behavioural patterns \cite{Fowler2009Cooperative}. Moreover, cascades may appear in both online~\cite{Leskovec2007Patterns,Duan2009Informational,Bond2012Million,Hui2012Information,Hodas2014Simple} and offline~\cite{Green2017Modeling} social environments.

\begin{figure}[!ht]
\centering
\includegraphics[width=1.\textwidth]{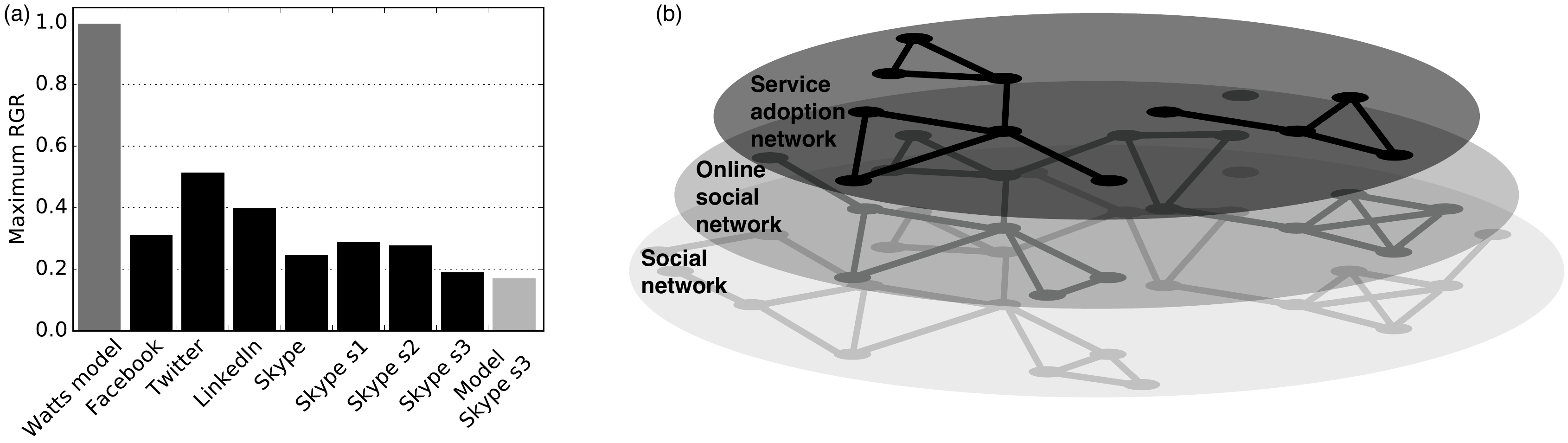}
\caption{\textbf{The speed and layers of online service adoption.} \textbf{(a)} Yearly maximum relative growth rate (RGR) of cumulative adoptions obtained by taking the maximum of the yearly adoption rate (yearly count of adoptions) normalised by the final observed number of adoptions of a given service. We show it for several online social-communication services \cite{SocialMedia} (black bars), including three paid Skype services (s1 - "subscription", s2  - "voicemail", and s3 - "buy credit"). The dark grey bar corresponds to a rapid cascade of adoption as suggested by the Watts threshold model, while the light grey bar is the prediction of our model for Skype s3. \textbf{(b)} Schematic layer structure of online service adoption systems. The lowest layer represents a real, offline social network; the middle layer corresponds to any online social network; and the top layer is the adoption of a service within the social network. As an advantage in this study we have full knowledge about the Skype online social network in this multi-layer structure, while we follow a paid service spreading on the online network.}
\label{fig:0}
\end{figure}

The characterisation \cite{Goel2012Structure,BorgeHolthoefer2013Cascading,Hackett2013Cascades,Gleeson2008Cascades,Brummitt2011,GhoshCascadesArxiv2010} and modelling \cite{Watts2002Simple,Hurd2013Watts,Singh2013Thresholdlimited,Gleeson2007Seed,Gleeson2014Simple} of behavioural cascades have received a lot of attention in the past and provide some understanding of the causal mechanisms and structure of empirical and synthetic cascades on various types of networks~\cite{Yagan2012Analysis,Brummitt2015Cascades,Karimi2013Threshold, Backlund2014Effects}. However, these studies fail in addressing the temporal dynamics of the emerging cascades, which may vary among empirical examples of social contagion. In other words, previous works do not answer why real-world cascades may evolve either slowly or rapidly over time. In contrast to the cases of rapid cascading mentioned above, the propagation of products in social networks is typically slower, with adoption spreading gradually, even if it is driven by threshold mechanisms and may eventually cover a large fraction of the total population~\cite{Karsai2014Complex}. This slow behaviour characterises the adoption of online services such as Facebook, Twitter, LinkedIn and Skype (Fig.\ref{fig:0}a), since their yearly maximum relative growth rate of cumulative adoption~\cite{SocialMedia} is lower than in the case of rapid cascades, as suggested in standard models of threshold-driven social contagion like the Watts threshold (WT) model~\cite{Watts2002Simple}.

In this chapter we review recent works~\cite{Ruan2015,Karsai2016Local} focusing on the empirical characterisation and mathematical modelling of the slow, threshold-driven spreading of service adoption in online social networks, particularly in the case of Skype. We first provide empirical evidence of the distribution of individual adoption thresholds and other structural and dynamical features of the worldwide Skype adoption cluster. We then show how to incorporate the observed structural and threshold heterogeneities into a dynamical threshold model where multiple individuals may adopt spontaneously (i.e. firstly among their acquaintances). We find that if the fraction of users who reject to adopt a product or idea in the model is large, the system enters a quenched state where the evolution and structure of the global adoption cluster is very similar to our observations of services within Skype. Model calculations and the analysis of the real social contagion process suggest that the evolving structure of an adoption cluster differs radically from previous expectations~\cite{Watts2002Simple}, since it is triggered by several spontaneous adoptions arriving at a constant rate. Furthermore, the stable adopters (who initially resist exposure) are actually responsible for the emergence of global social adoption.

\section{Empirical observations}
\label{sec:emp}

In order to observe service adoption dynamics we analyse an example of an online diffusion process, where we have access to individual service adoption events as well as the underlying social network. Our aim is to identify the crucial mechanisms necessary to consider in models of complex contagion to match them better with reality, and define a model that incorporates these mechanisms and captures the possible dynamics leading to the emergence of real-world global cascades. 

To fully understand service adoption processes on online social structures, we need to keep in mind some of their proxy characteristics. People of a society constitute a social network by being connected with ties of several kinds that are maintained in various ways. However, and despite their recent popularity, online social systems are not capable of mapping the entire social network as offline, occasionally maintained, temporary, or ill-favored social ties may remain invisible in such systems. Therefore, these networks provide only a proxy sample of the real social structure (Fig.\ref{fig:0}b), with important but also insignificant social ties present. Moreover, data available for social network studies commonly arrives as a sample of a larger online social system, which unavoidably leads to observational biases. In addition, connections in an online social structure cannot precisely assign the flow of direct social influence among the connected individuals, only the possibility of it. Finally, just like real social networks, online social systems evolve over time via the creation and dissolution of social ties or by nodes entering or leaving the system. Due to all these limitations it is rather challenging to make unbiased observations about any unfolding dynamical processes, without making some assumption about the underlying online social systems.

In our study we use the social network of one of the largest voice-over-internet providers in the world, the network of Skype, which actually copes well with the limitations listed above. It maps all connections in the Skype network without sampling, thus it provides us with a complete, unbiased map of the underlying social network, maintaining the diffusion of services available only for registered users in the network. This network evolves as a function of time via adoption, churning, and link creation dynamics. We have shown in an earlier study~\cite{Karsai2014Complex} that while rates of these actions increase considerably with time, the adoption processes can be well characterised by the net rate of the actual number of users. We also found that while spontaneous adoptions and churning evolve with a constant rate, the probability of peer-pressured adoptions corresponds linearly to the strength of social influence, giving rise to a non-linear dynamics at the system level, which enables its modelling as a complex contagion process. 

In our study we concentrate on the adoption dynamics of a paid service that unfolds over the Skype social network (Fig.\ref{fig:0}b). Since this adoption process evolves in a considerably faster time-scale than the underpinning social network, we can validly assume a time-scale separation. Thus, from here on we consider the network structure to be static, which may give us a good first approximation while concentrating on the adoption dynamics unfolding on its fabric. To identify the effects of social influence in our empirical system we also present a null model study (Section~\ref{sec:sinf}).

\subsection{Data description}

In our social network nodes represent users and edges between pairs of users exist if they are in each other's contact lists. A user's contact list is composed of \emph{friends}. If user $u$ wants to add another user $v$ to his/her contact list, $u$ sends $v$ a contact request, and the edge is established at the moment $v$ approves the request (or not, if the contact request is rejected). For the purpose of our study we use the largest connected component of the aggregated free Skype service network, which was recorded from September 2003 to November 2011 (i.e. over $99$ months) and contains roughly 4.4 billion links and 510 million registered users worldwide~\cite{SkypeIPO}. The data is fully anonymised and considers only confirmed connections between users after the removal of spammers and blocked nodes.

To study an example of service adoption dynamics we follow the purchases of the ``buy credit'' paid service for $89$ months starting from 2004. Data includes the time of first payment of each adopting user, an individual and conscious action that tracks adoption behaviour. Note that other examples about the adoption dynamics of similar services are presented in \cite{Karsai2016Local}.

\subsection{Degree and threshold heterogeneities}

In his seminal work on modelling adoption cascades~\cite{Watts2002Simple}, Watts identified two structural characteristics that control the emergence of collective adoption cascades. One is the distribution $P(k)$ of degrees (i.e. number of neighbours of a node), with average $z = \langle k \rangle$, and the other is the distribution $P(\varphi)$ of adoption thresholds (with average $w = \langle \varphi \rangle$), defined as the necessary fraction of exposed neighbours that triggers the adoption of an individual under study, or central ego. 

\begin{center}
\begin{figure*}
\includegraphics[width=\textwidth,angle=0]{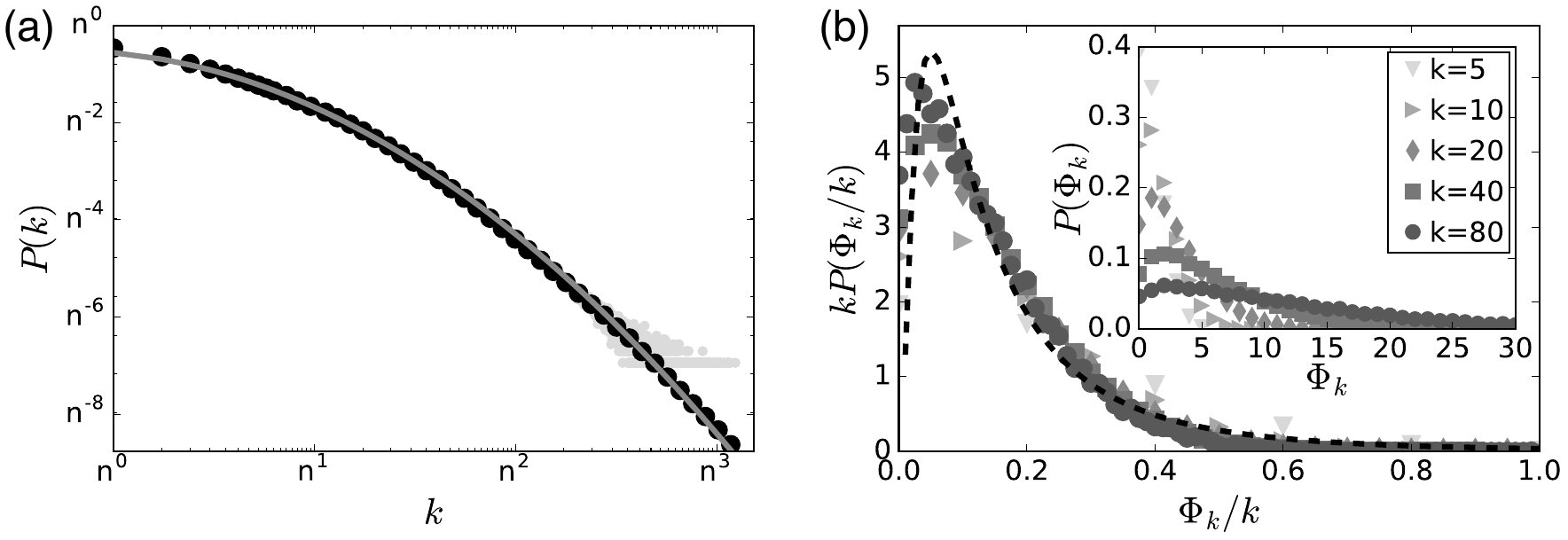}
\caption{\textbf{Degree and threshold heterogeneities.} \textbf{(a)} Degree distribution $P(k)$ of the Skype network (light/dark grey circles for raw/binned data) on a double log-scale with arbitrary base $n$. $P(k)$ is fitted with a log-normal distribution (see text) with parameters $\mu_D=1.2$ and $\sigma_D=1.39$, and average $z = 8.56$ (gray line). \textbf{(b)} Distribution $P(\Phi_k)$ of integer thresholds $\Phi_k$ for several degree groups in Skype s3 (inset). By using $P(\Phi_k, k) = k P(\Phi_k/k)$, these curves collapse into 
a master curve approximated by a log-normal function (dashed line in main panel) with parameters $\mu_T=-2$ and $\sigma_T=1$, as constrained by the average threshold $w = 0.19$.
\label{fig:1}}
\end{figure*}
\end{center}

Degree heterogeneities have been in the focus of network science for a while now, and a broad degree distribution $P(k)$ is one of the main characteristics of complex networks~\cite{Newman2010Networks,Barabasi2016NetSci}. This distribution has been usually described as a power-law, but a log-normal fit has often turned out to work better~\cite{Mitzenmacher2004}. The latter is the case with our data:
\begin{equation}
P(k) \propto k^{-1} \exp[-(\ln k - \mu_D)^2/(2\sigma_D^2)],
\label{eq:Pk}
\end{equation}
where the best fit is obtained with $k \geq \kmin$ and parameters $\mu_D=1.2$, $\sigma_D=1.39$ and $\kmin=1$ (Fig.~\ref{fig:1}a), giving an average degree $z = 8.56$. 

It is a challenging task to quantify individual adoption thresholds, as their observation simultaneously requires information about the underlying network structure and the dynamical adoption process evolving on top. Therefore, beside measuring the number $k$ of friends of an ego in the Skype social network (already needed for the degree distribution), for $k$-degree users at the time of their adoption we measure the number $\Phi_k$ of their neighbours who have adopted the service earlier, i.e. the integer threshold~\cite{Gleeson2008Cascades}. To our knowledge, this is the first detailed study measuring the number of adopting neighbors of adopters in an empirical setting. The obtained distribution $P(\Phi_k)$ for varying $k$ is shown in the inset of Fig.~\ref{fig:1}b. The importance of our empirical findings is amplified by the observation that these distributions can be scaled together when using the fractional threshold variable $\varphi=\Phi_k/k$, i.e. the fraction of adopting neighbors at the time of adoption (Fig.~\ref{fig:1}b main panel). Thus, in a discussion of whether the number or the ratio of adopting neighbours matters in behavioural adoption~\cite{Watts2002Simple,Centola2007Complex}, our results give strong support to the latter.

Using fractional thresholds and the relationship $P(\Phi_k, k) = k P(\Phi_k/k)$, all distributions collapse to a master curve, which is once again well-approximated by a log-normal function of the following form,
\begin{equation}
P(\Phi_k/k)=P(\varphi) \propto \varphi^{-1} \exp[-(\ln\varphi - \mu_T)^2/(2\sigma_T^2],
\label{eq:Pphi}
\end{equation}
with parameters $\mu_T=-2$ and $\sigma_T=1$ as constrained by the average threshold $w = 0.19$~\cite{Karsai2016Local}. 
These empirical observations, in addition to the broad degree distribution, provide quantitative description of the heterogeneous nature of adoption thresholds.

\subsection{Dynamics and structure of adoption cascades}

\begin{figure} \centering
\includegraphics[width=\textwidth]{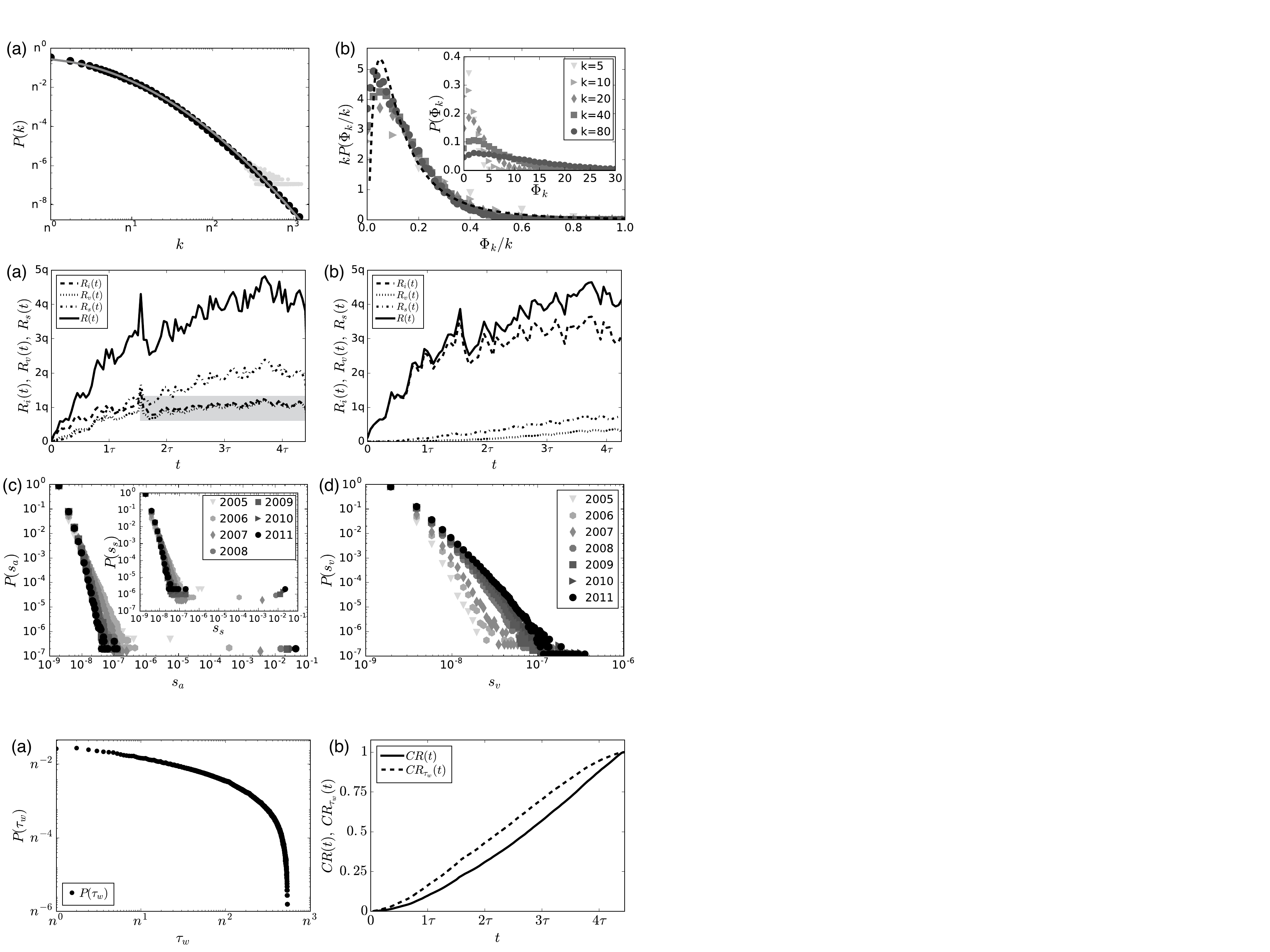}
\caption{\small{\bf The dynamics and structure of adoption cascades.} \textbf{(a)} Adoption rate of innovators [$R_i(t)$], vulnerable nodes [$R_v(t)$], and stable nodes [$R_s(t)$], as well as the net service adoption rate [$R(t)$], where the rates 
are measured with a 1-month time window, and $q$ and $\tau$ are arbitrary constants. The shaded area indicates the regime where innovators adopt approximately with constant rate. {\bf (b)} Null model rates where times of adoption are randomly shuffled. \textbf{(c)} Empirical connected-component size distribution at different times for the adoption [$P(s_a)$, main panel] and stable adoption [$P(s_s)$, inset] networks, with $s_a$ and $s_s$ relative to system size. \textbf{(d)} Empirical connected-component size distribution $P(s_v)$ for the relative size of innovator-induced vulnerable trees at different times.}
\label{fig:2}
\end{figure}

Since we know the complete structure of the online social network, as well as the first time of service usage for all adopters, we can follow the temporal evolution of the adoption dynamics. By counting the number of adopting neighbours of an ego, we identify innovators ($\Phi_k=0$), and vulnerable ($\Phi_k=1$) or stable ($\Phi_k>1$) nodes, in accordance with the categorisation of Watts \cite{Watts2002Simple}. As we show in Fig.~\ref{fig:2}a, the adoption rates for these categories behave rather differently from previous suggestions \cite{Watts2002Simple}. First, there is not only one seed but an increasing fraction of innovators in the system who, after an initial period, adopt approximately at a constant rate (denoted by the grey shaded area in Fig.~\ref{fig:2}a). Second, vulnerable nodes adopt approximately with the same rate as innovators, which suggests a strong correlation between these types of adoption. This stationary behaviour is rather surprising as environmental effects, like competition or marketing campaigns, could potentially influence the adoption dynamics. On the other hand, the overall adoption process accelerates due to the increasing rate of stable adoptions induced by social influence. 

To better understand how innovation spreads throughout the social network, we take a closer look at the internal structure of the service adoption process. To do so, we consider individual adoption times and construct an evolving adoption network, where links exist between users who have adopted the service before time $t$ and are connected in the social network underneath. In order to avoid the effect of instantaneous group adoptions (evidently not driven by social influence), we only consider links between connected nodes whose adoption did not happen at the same time. This way links in the adoption graph indicate ties where social influence among individuals could have existed. By observing the evolution of the adoption network, we are interested in its connectedness and its composition of sub-components of adopters of different kinds.

The size distribution $P(s_a)$ of connected components in the adoption network shows the emergence of a giant 
percolating component over time (Fig.~\ref{fig:2}c main panel), along with several other small clusters. Moreover, after decomposition we observe that the giant cluster builds up from several innovator seeds that induce small vulnerable trees locally (Fig.~\ref{fig:2}d), each with small depth \cite{Karsai2016Local,Bakshy11,Goel2012Structure}. At the same time the stable adoption network (considering connections between all stable adopters at the time) has a giant connected component, indicating the emergence of a percolating stable cluster with size comparable to the largest adoption cluster (Fig.~\ref{fig:2}c, inset). These observations suggest a scenario for the evolution of the global adoption component where multiple innovators adopt at different times and trigger local vulnerable trees, which in turn induce a percolating component of the connected stable nodes holding 
the global adoption cluster together. Consequently, in the structure of the adoption network primary triggering effects are important only locally, while external and secondary triggering mechanisms seem to be responsible for the emergence of global-scale adoption. 

Despite of this expansion dynamics and connected structure of the service adoption network, we need to take a closer look at spurious effects, which could potentially induce the observed behaviour. First, during our analysis we assume that the adoption process is exclusively driven by social influence, without any direct information about the presence of the influence itself. One can argue that the observed phenomena is simply explained by homophily, i.e. by frequent links between people who are both interested in the given service and who would adopt independently from each other. Second, the service reaches less than $6\%$ of the total number of active Skype users over a period of $7$ years~\cite{SkypeIPO}. Since this adopting minority is connected within a giant adopting cluster, it may indicate local effects of social influence but also raises the question about the role of non-adopting users. Finally, we observe that the giant adoption cluster evolves over several years, which could simply be the consequence of individual decisions of users to wait to adopt the service even after their threshold has been reached. In the following we further investigate these questions to better understand the adoption process. First we present a null model study to underline the overall effects of social influence as compared to homophily; we also perform a time re-scaling experiment to explore the role of waiting times on the global adoption dynamics; and finally we propose a dynamical threshold model \cite{Ruan2015,Karsai2016Local}, which helps us understand the role of multiple innovators and non-adopters in the unfolding of the service adoption processes.

\subsection{Social influence vs. homophily}
\label{sec:sinf}

Studies of social contagion phenomena assume that social influence is responsible for the correlated adoption of connected people. However, an alternative explanation for the observed correlated adoption patterns is homophily: a link creation mechanism by which similar egos get connected in a social structure. In the latter case, the correlated adoption of a connected group of people would be explained by their similarity and not necessarily due to social influence. Homophily and influence are two processes that may simultaneously play a role during the adoption process. Nevertheless, distinguishing between them on the individual level is very challenging using our or any similar datasets \cite{Shalizi2011,Aral2009}. Fortunately, at the system level one may identify which process is dominant in the empirical data. To do that we first need to elaborate on the differences between these two processes.

Influence-driven adoption of an ego may take place once one or more of its neighbours have adopted, since then their actions may influence the decision of the central ego. Consequently, the time ordering of adoptions of the ego and its neighbours matters. Homophily-driven adoption is, however, different. Homophily drives social tie formation such that similar people tend to be connected in the social structure. In this case connected people may adopt because they have similar interests, but the time ordering of their adoptions would not matter. Therefore, it is valid to assume that adoption could evolve in clusters due to homophily, but these adoptions would appear in a more-or-less random order.

To test this hypothesis we define a null model where we take the adoption times of users and shuffle them randomly among all adopting egos. This way a randomly selected time is assigned to each adopter, while the adoption rate and the final set of adopters remain the same. Moreover, this procedure only destroys correlations between adoption events induced by social influence, but keeps the social network structure and node degrees unchanged. In this way, during the null model process the same egos appear as adopters, but the rates of adoption may in principle change (or not), corresponding to social influence (or homophily) as a dominant factor during the adoption process. If adoption is mostly driven by homophily, the rates of adoption would not change considerably beyond statistical fluctuations. On the other hand, if social influence plays a role in the process, rates of adoption in the null model should be very different from the empirical curves, implying that the time ordering of events matters in the adoption process. In this case, the rate of innovators should be higher than in the empirical data, since nodes that are in the adoption cluster originally without being directly connected, would have a greater chance to appear as innovators, due to a random adoption time that is not conditional to the time ordering of the adopting neighbours.

After calculating the adoption rates of different user groups in the shuffled null model, we observe the latter situation (Fig.~\ref{fig:2}b): the rate of innovators becomes dominant, while the rates of stable and vulnerable adoptions drop considerably as they appear only by chance. This suggests that the temporal ordering of adoption events matters a lot in the evolution of the observed adoption patterns, and thus social influence may play a strong role here. Of course one cannot decide whether influence is solely driving the process or homophily has some impact on it; in reality it probably does to some extent. However, we can use this null model measure to demonstrate the presence and importance of the mechanism of influence during the adoption process.

\subsection{Waiting time of adoption}
\label{sec:tw}

\begin{figure}[!ht] \centering
\includegraphics[width=\textwidth]{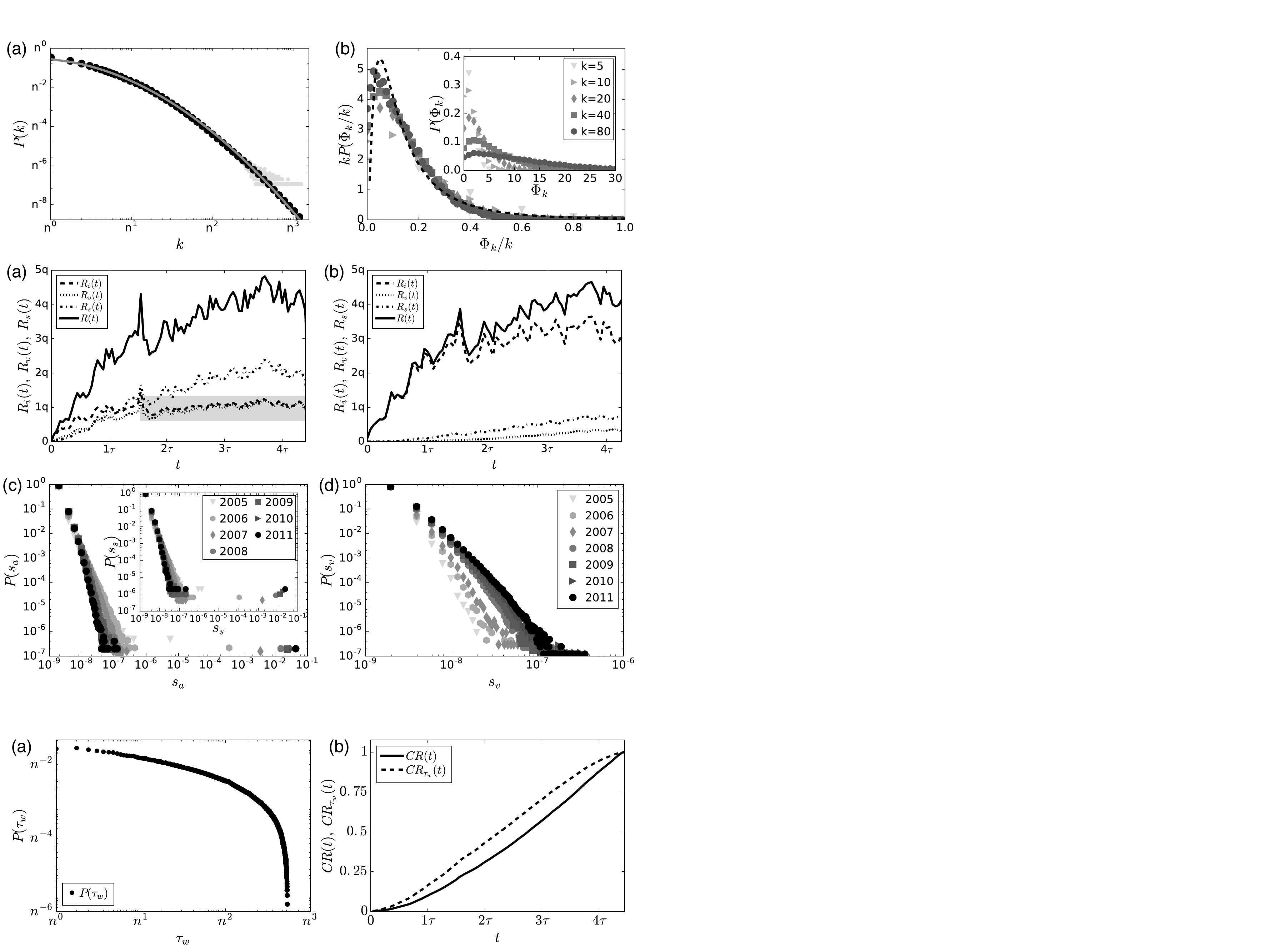}
\caption{\small{\bf The waiting time distribution and its effect on the adoption process.} {\bf (a)} Distribution $P(\tau_w)$ of times between the last adoption in the egocentric network of an individual and his/her own adoption. {\bf (b)} Cumulative adoption rates before and after the removal of waiting times [$CR(t)$ and $CR_{\tau_w}(t)$, respectively]. $n$ and $\tau$ are arbitrary constant values.}
\label{fig:3}
\end{figure}

As we mentioned earlier, one reason behind the slow evolution of the adoption process could be due to the time users wait after their personal adoption threshold is reached and before adopting the service. This lag in adoption can be due to individual characteristics, or can come from the fact that social influence does not spread instantaneously (as commonly assumed in threshold models). This waiting time $\tau_w$ can be estimated by measuring the time difference between the last adoption in a user's egocentric network and the time of his/her adoption. This time is $\tau_w=0$ by definition for innovators, but $\tau_w$ can take any positive value for vulnerable and stable adopters up to the length of the observation period.

We find that waiting times are broadly distributed for adopters in our dataset (Fig.~\ref{fig:3}a), meaning that many users adopt the service shortly after their personal threshold is reached, but a considerable fraction waits long before adopting the service. This heterogeneous nature of waiting times may be a key element behind the observed adoption dynamics. One way to figure out its effect on the speed of cascade evolution is by removing them. We can extract the waiting time from the adoption time of adopters and assign a rescaled adoption time for each of them. The rescaled adoption time of a user is the last time when his/her fraction of adopting neighbours changed and the adoption threshold was (hypothetically) reached. After this procedure, we can calculate a new adoption rate function by using the rescaled adoption times and compare this rate to the original. From Fig.~\ref{fig:3}b we conclude that although adoption becomes faster, the rescaled adoption dynamics is still not rapid. On the contrary, it suggests that the rescaled adoption dynamics is still very slow and quite similar to the original. Consequently, waiting times cannot explain the observed slow dynamics of adoption.

Note that long waiting times can have a further effect on the measured dynamics. After the `real' threshold of a user is reached and he/she waits to adopt, some neighbours may adopt the product. Hence all observed measures are in this sense `effective': observed thresholds are larger or equal than real thresholds; the innovator rate is smaller or equal; the vulnerable and stable rates will be larger or equal; and waiting times will be shorter or equal than the real values. Consequently the process may actually be 
faster than that we observe in Fig.~\ref{fig:3}b after removing the effective waiting times. However, this bias becomes important only after the majority of individuals in the social network has adopted the service and the spontaneous emergence of adopting neighbours becomes more frequent. As the fraction of adopters in our dataset is always less than $6\%$ \cite{SkypeIPO}, we expect minor effects of this observational bias on our measurements.

\section{Modelling social contagion}
\label{sec:model}

In order to understand better the possible microscopic mechanisms behind the empirical observations of online service adoption described previously, we introduce and analyse two agent-based network models of threshold-driven social contagion. First we discuss the WT model as originally proposed by Watts~\cite{Watts2002Simple}, and secondly an extended, dynamical threshold model devised by us~\cite{Ruan2015,Karsai2016Local}, where both multiple innovators and non-adopters have a role in social contagion.

\subsection{The Watts model}
\label{sec:wmodel}

Under the complex contagion hypothesis by Granovetter, Centola and others~\cite{Granovetter1978Threshold,Centola2007Complex}, social contagion may be modelled as a binary-state process evolving in 
a network and driven by a threshold mechanism. In this framework individuals are represented by agents or network nodes, each in either a susceptible (0) or adopter (1) state, while the influence by an agent is achieved by transferring information via social ties. Nodes are connected in a network with degree distribution $P(k)$ and average degree $z = \langle k \rangle$. Moreover, each node has an individual threshold $\varphi \in [0, 1]$ drawn from a distribution $P(\varphi)$ with average $w = \langle \varphi \rangle$. The threshold $\varphi$ determines the minimum fraction of exposed neighbours that triggers adoption, capturing the resistance of an individual against engaging in a given behaviour. Hence, in case  
a node has $m$ adopting neighbours and $m \geq k \varphi$ (the so-called {\it threshold rule}), it switches state from $0$ to $1$ and remains so until the end of the dynamics. In his seminal paper about threshold dynamics~\cite{Watts2002Simple}, Watts classified nodes into three categories based on their threshold and degree: He first identified {\it innovator} nodes that spontaneously change state to $1$ and therefore start the spreading process. Such nodes have a trivial threshold $\varphi=0$. Then there are nodes with threshold $0 < \varphi \leq 1/k$, called {\it vulnerable}, which need one adopting neighbour before their own adoption. Finally, there are more resilient nodes with threshold $\varphi>1/k$, known as {\it stable}, representing individuals in need of strong social influence to follow the actions of their acquaintances.

In the WT model~\cite{Watts2002Simple}, small perturbations (like the spontaneous adoption of a single seed node) can trigger network-wide cascading patterns.  However, their emergence is subject to the following {\it cascade condition}: the innovator seed has to be linked to a percolating vulnerable cluster, which adopts immediately afterwards and further triggers a {\it global} cascade (i.e. a set of adopters larger than a fixed fraction of a finite network, or a nonzero fraction of adopters in an infinite network). The cascade condition is satisfied if the network is inside a bounded regime in $(w, z)$-space \cite{Watts2002Simple}. When considering a vanishingly small innovator seed and a configuration-model network~\cite{Newman2010Networks} [i.e., by ignoring structural correlations in the social network and characterising it solely by its degree distribution $P(k)$], a generating function approach allows us to write the cascade condition as 
\begin{equation}
\label{eq:wattsCond}
\sum_k \frac{k}{z} (k - 1) P(k) f(k, 1) > 1,
\end{equation}
where $f(k, 1) = C(1/k)$ is the probability that a randomly-selected node with degree $k$ is vulnerable, and $C$ is the cumulative distribution function of $P(\varphi)$. More generally, $f(k, m)$ (for $m = 0,\ldots,k$) is also known as a response function of the monotone binary dynamics defining the WT model~\cite{porter2016,Ruan2015}.

As Eq.(\ref{eq:wattsCond}) shows, the cascade regime depends on degree and threshold heterogeneities~\cite{Watts2002Simple} and may change its shape if several innovators start the process~\cite{Singh2013Thresholdlimited}. In addition, while models with more sophisticated functional forms of social influence may be introduced \cite{Latane1981The, Dodds2004Universal}, the original assumption proposed by Watts and Granovetter seems to be sufficient to interpret our observations.

\subsection{Dynamical threshold model with immune nodes}
\label{sec:dmodel}

\begin{figure}
\includegraphics[width=\linewidth]{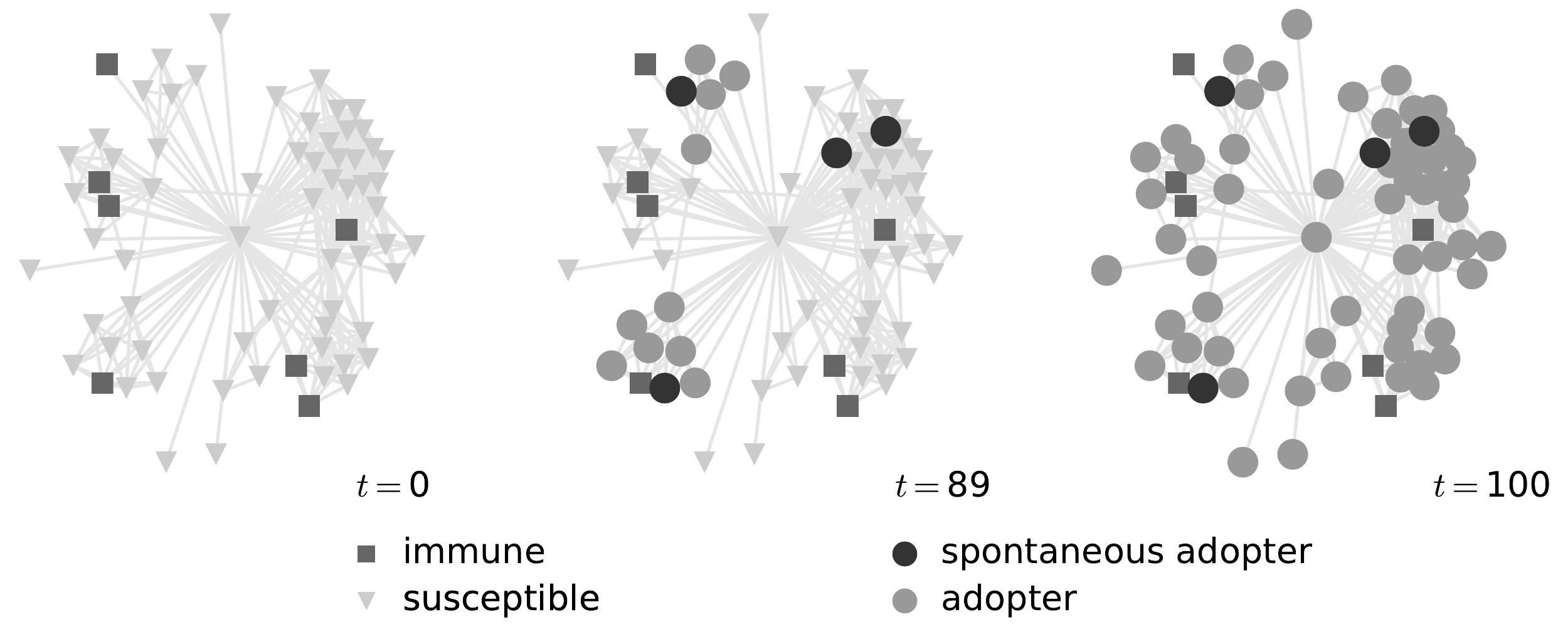}
\caption{{\bf Immune individuals in social contagion.} Numerical simulation of our dynamical threshold model in an empirical network, with a single adoption threshold $\varphi = 0.2$ for all the nodes, rate of spontaneous adopters $\pn = 0.0005$, and fraction of immune nodes $r = 0.1$. The network is an ego sample of Facebook friendships with size $N = 96$ and average degree $z = 10.63$~\cite{Leskovec2012Learning}. The network shows how susceptible nodes adopt spontaneously with rate $\pn$, or after a fraction $\varphi$ of their neighbours has adopted, while immune nodes never adopt.}
\label{Fig:4}
\end{figure}

Our modelling framework is an extension to the WT model and similar threshold dynamics on networks, studied by Watts, Gleeson, Singh, and others, where all the nodes are initially susceptible and innovators are only introduced as an initial seed of arbitrary size~\cite{Watts2002Simple,Karampourniotis2015The,Gleeson2007Seed,Singh2013Thresholdlimited}. Apart from the above discussed threshold rule and motivated by the empirical observations in the spread of online services within Skype, our model considers two additional features, namely that (i) a fraction $r$ of `immune' nodes never adopts, indicating a lack of interest in the online service, and that (ii) due to external influence, susceptible nodes adopt the service spontaneously (i.e. become innovators) throughout the time with constant rate $\pn$, rather than only at the beginning of the dynamics. In this way, the dynamical evolution of the system is completely determined by the online social network, the distribution $P(\varphi)$ of thresholds, and the parameters $r$ and $\pn$ (Fig.~\ref{Fig:4}). For the sake of simplicity, we consider a configuration-model network and statistical independence between degrees and thresholds~\cite{Gleeson2008Cascades,gleeson2013binary,gleeson2011high}. We remark that the somewhat similar concepts of `stubborn nodes', mimicking individuals' resistance against adoption~\cite{Brummitt2012Multiplexity,Lee2014Threshold}, and `global nodes', capturing adoption driven by external effects~\cite{Kobayashi2015Trend}, have also been considered in threshold models and show a rich variety of effects on cascading behaviour.

Our threshold model~\cite{Ruan2015,Karsai2016Local} can be studied analytically by extending the framework of approximate master equations (AMEs) for monotone binary-state dynamics, as recently developed by Gleeson~\cite{Gleeson2008Cascades,gleeson2013binary,gleeson2011high}, where the transition rate between susceptible and adoption states only depends on the number $m$ of network neighbours that have already adopted. We describe a node by the property vector $\kvec = (k, c)$, where $k = k_0, k_1, \ldots k_{M-1}$ is its degree and $c = 0, 1, \ldots, M$ its type, i.e. $c = 0$ is the type of the fraction $r$ of immune nodes, while $c \neq 0$ is the type of all non-immune nodes that have threshold $\varphi_c$. In this way, $P(\varphi)$ is substituted by the discrete distribution of types $P(c)$ (for $c > 0$). The integer $M$ is the maximum number of degrees (or non-zero types) considered in the AME framework, which can be increased to improve the accuracy of the analytical approximation at the expense of speed in its numerical computation.

We characterise the static social network by the extended distribution $\Pk$, where $\Pk = r P(k)$ for $c = 0$ and $\Pk = (1 - r) P(k) P(c)$ for $c > 0$. Non-immune and susceptible nodes with property vector $\kvec$ adopt spontaneously with a constant rate $\pn$, otherwise they adopt only if a fraction $\varphi_c$ of their $k$ neighbours has adopted before. These rules are condensed into the probability $\Fkm dt$ that a node will adopt within a small time interval $dt$, given that $m$ of its neighbours are already adopters,
\begin{equation}
\label{eq:thresRule}
\Fkm =
\begin{cases}
\pr & \text{if} \quad m < k \varphi_c \\
1 & \text{if} \quad m \geq k \varphi_c
\end{cases}, \quad \forall m \; \text{and} \; k, c \neq 0,
\end{equation}
with $F_{(k,0),m} = 0$ $\forall k, m$ and $F_{(0,c),0} = \pr$ $\forall c \neq 0$ (for immune and isolated nodes, respectively). The rescaled rate $\pr = \pn / (1 - r)$ (with $\pr = 1$ for $\pn > 1 - r$) is necessary if we wish to obtain a rate $\pn$ of innovators for early times of the dynamics, regardless of the value of $r$.

The dynamics of adoption is well described by an AME for the fraction $\skm(t)$ of $\kvec$-nodes that are susceptible at time $t$ and have $m=0,\ldots,k$ adopting neighbours~\cite{porter2016,gleeson2013binary,gleeson2011high},
\begin{equation}
\label{eq:AMEsThres}
\dskm = -\Fkm \skm -\bs (k - m) \skm + \bs (k - m + 1) \skmo,
\end{equation}
where
\begin{equation}
\label{eq:rateBs}
\bs(t) = \frac{\sumk \Pk \summ (k - m) \Fkm \skm(t)}{\sumk \Pk \summ (k - m) \skm(t)},
\end{equation}
and the sum is over all the degrees and types, i.e. $\sumk \bullet = \sum_k \sum_c \bullet$. To reduce the dimensionality of Eq.~(\ref{eq:AMEsThres}), we consider the ansatz
\begin{equation}
\label{eq:AMEansatz}
\skm(t) = \Bkm [\nu(t)] e^{-\pr t}
\quad \text{for} \; m < k\varphi_c  \; \text{and} \; c \neq 0,
\end{equation}
with $\nu(t)$ the probability that a randomly-chosen neighbour of a susceptible node is an adopter.

Introducing the ansatz of Eq.~(\ref{eq:AMEansatz}) into the AME system of Eq.~(\ref{eq:AMEsThres}) leads to the condition $\dot{\nu} = \bs (1 - \nu)$. With some algebra, the AMEs for our dynamical threshold model are reduced to the pair of ordinary differential equations
\begin{subequations}
\label{eq:reducedAMEs}
\begin{align}
\dot{\rho} &= h(\nu, t) - \rho, \\
\dot{\nu} &= g(\nu, t) - \nu,
\end{align}
\end{subequations}
where $\rho(t) =  1 - \sumk \Pk \summ \skm(t)$ is the fraction of adopters in the network, and the initial conditions are $\rho(0) = \nu(0) = 0$. Here,
\begin{equation}
\label{eq:hTerm}
h = (1 - r) \Big[ \ft + (1 - \ft) \sum_{\kvec | c \neq 0} P(k) P(c) \sum_{m \geq k\varphi_c} \Bkm(\nu) \Big],
\end{equation}
and
\begin{equation}
\label{eq:gTerm}
g = (1 - r) \Big[ \ft + (1 - \ft) \sum_{\kvec | c \neq 0} \frac{k}{z} P(k) P(c) \sum_{m \geq k\varphi_c} \Bkom(\nu) \Big],
\end{equation}
where $\ft = 1 - (1 - \pr) e^{-\pr t}$, and $\Bkm(\nu) = \binom{k}{m} \nu^m (1 - \nu)^{k - m}$ is the binomial distribution. The fraction of adopters $\rho$ is then obtained by solving Eq.~(\ref{eq:reducedAMEs}) numerically. Since the susceptible nodes adopt spontaneously with rate $p_n$, the fraction of innovators $\rho_0(t)$ in the network is given by
\begin{equation}
\label{eq:innovFrac}
\rho_0(t) = \pr \int_0^t [1 - r - \rho(\tau)] d\tau.
\end{equation}

\begin{center}
\begin{figure}[t]
\includegraphics[width=\textwidth,angle=0]{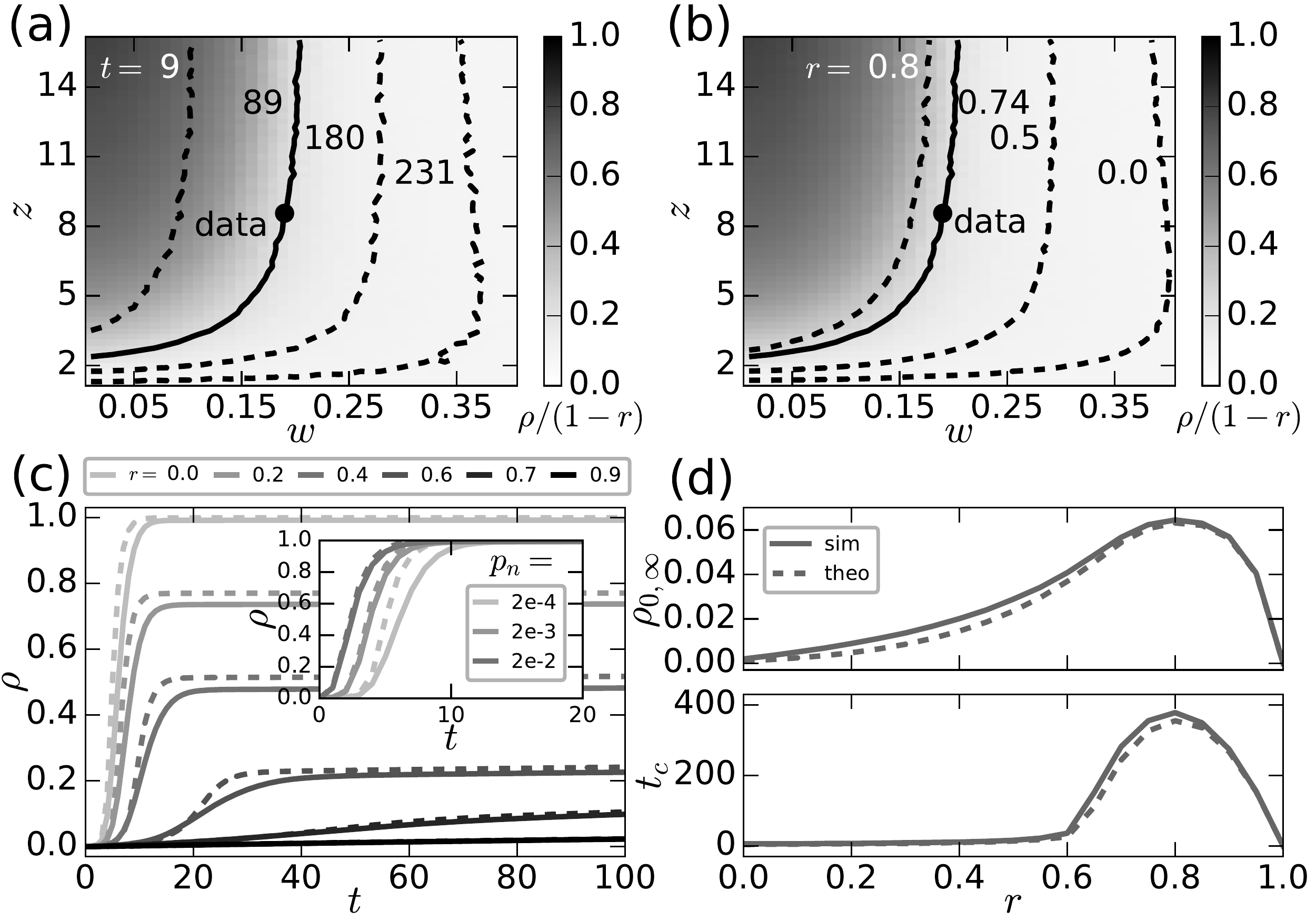}
\caption{{\bf A dynamical threshold model for the adoption of online services.} {\bf (a-b)} Surface plot of the normalised fraction of adopters $\rho / (1 - r)$ in $(w, z)$-space, for $r = 0.73$ and $t = 89$. Contour lines signal the parameter values for which $20\%$ of non-immune nodes have adopted, for fixed $r$ and varying time (a), and for fixed time and varying $r$ (b). The continuous contour line and dot indicate parameter values of the last observation of Skype s3. A regime of maximal adoption ($\rho \approx 1 - r$) grows as time goes by, and shrinks for larger $r$. {\bf (c)} Time series of the fraction of adopters $\rho$ for fixed $p_n = 0.00019$ and varying $r$ (main), and for fixed $r = 0$ and varying $p_n$ (inset). These curves are well approximated by the solution of Eq.~(\ref{eq:reducedAMEs}) for $k_0 = 3$, $k_{M-1} = 150$ and $M = 25$ (dashed lines). The dynamics is clearly faster for larger $p_n$ values. As $r$ increases, the system enters a regime where the dynamics is slowed down and adopters are mostly innovators. {\bf (d)} Final fraction of innovators $\rho_{0,\infty}$ and the time $t_c$ when $50\%$ of non-immune nodes have adopted as a function of $r$, both simulated and theoretical. The crossover to a regime of slow adoption is characterised by a maximal fraction of innovators and time $t_c$. Unless otherwise stated, $p_n=0.00019$ and we use $N=10^4$, $\mu_D=1.09$, $\sigma_D=1.39$, $\kmin=1$, $\mu_T=-2$, and $\sigma_T=1$ to obtain $z = 8.56$ and $w = 0.19$ as in Skype s3. The difference in $\mu_D$ between data and model is due to finite-size effects. Numerical results are averaged over $10^2$ (a-b) and $10^3$ (c-d) realisations.
\label{fig:5}}
\end{figure}
\end{center}

We may also implement our dynamical threshold model numerically via a Monte Carlo simulation in a network of size $N$, with a log-normal degree distribution and a log-normal threshold distribution as observed empirically in the case of Skype. Hence we can explore the behaviour of $\rho$ and $\rho_0$ as a function of $z$, $w$, $p_n$ and $r$, both in the numerical simulation and in the theoretical approximation given by Eqs.~(\ref{eq:reducedAMEs}) and~(\ref{eq:innovFrac}). For $p_n > 0$ some nodes adopt spontaneously as time passes by, leading to a frozen state characterised by the final fraction adopters $\rho(\infty) = 1 - r$. However, the time needed to reach such a state depends heavily on the distribution of degrees and thresholds, as indicated by a region of large adoption ($\rho \approx 1 - r$) that grows in $(w, z)$-space with time (contour lines in Fig.~\ref{fig:5}a). If we fix the time in the dynamics and vary the fraction of immune nodes instead, this region shrinks as $r$ increases (contour lines in Fig.~\ref{fig:5}b). In other words, the set of networks (defined by their average degree and threshold) that allow the spread of adoption is larger at later times in the dynamics, or when the fraction of immune nodes is small. When both $t$ and $r$ are fixed, the normalised fraction of adopters $\rho / (1 - r)$ gradually decreases for less connected networks with larger thresholds (surface plot in Fig.~\ref{fig:5}a and b).

Both numerical simulations and analytical approximations show how the dynamics of spreading changes by introducing immune individuals in the social network. For $r \approx 0$, the adoption cascade appears sooner for larger $\pn$, since this parameter regulates how quickly we reach the critical fraction of innovators necessary to trigger a cascade of fast adoption throughout all susceptible nodes (Fig.~\ref{fig:5}c, inset). Yet as we increase $r$ above a critical value $r_c$ (and thus introduce random quenching), the system enters a regime where rapid cascades disappear and adoption is slowed down, since stable nodes have more immune neighbours and it is difficult to fulfil their threshold condition. The crossover between these fast and slow regimes is gradual, as seen in the shape of $\rho$ for increasing $r$ (Fig.~\ref{fig:5}c, main panel). We may identify $r_c$ in various ways: by the maximum in both the final fraction of innovators $\rho_{0,\infty} = \rho_0(\infty)$ and the critical time $t_c$ when $\rho = (1-r)/2$ (Fig.~\ref{fig:5}d), or as the $r$ value where the inflection point in $\rho$ disappears. These measures indicate $r_c \approx 0.8$ for parameter values calibrated with Skype data. All global properties of the dynamics (like the functional dependence of $\rho$ and $\rho_0$) are very well approximated by the solution of Eqs.~(\ref{eq:reducedAMEs}) and~(\ref{eq:innovFrac}) (dashed lines in Fig.~\ref{fig:5}c and d). Indeed, the AME framework is able to capture the shape of the $\rho$ time series, the crossover between regimes of fast and slow adoption, as well as the maximum in $\rho_{0,\infty}$ and $t_c$.

In the simplified case of an Erd\H{o}s-R\'{e}nyi random graph as the underlying social network, the crossover between fast and slow regimes of spreading may also be characterised by a percolation-type transition in the asymptotic limit ($t \to \infty$) of the size distribution $P(s)$ of {\it induced} adoption clusters, i.e. connected components of adopters disregarding innovators~\cite{Ruan2015}. For early times $P(s)$ includes small induced clusters only, which in turn indicates that a larger fraction of spontaneous adopters is crucial for global spreading in the absence of a percolating vulnerable component. However, for late times the behaviour of $P(s)$ differs between regimes: in the regime of fast spreading the distribution becomes bimodal due to the appearance of a global cluster of induced adopters, while in the slow regime it remains unimodal until the end of dynamics.

Finally, in the extreme case of $\pn = 0$ (corresponding to the WT model with immune nodes), the reduced AME system of Eq.~(\ref{eq:reducedAMEs}) can be used to derive a cascade condition and thus give insight into the dynamics of spreading in the presence of immune individuals~\cite{porter2016,Ruan2015}. Eq.~(\ref{eq:reducedAMEs}) has an equilibrium point for 
the initial condition $(\rho(0), \nu(0)) = (0, 0)$. If this equilibrium point is linearly unstable, the perturbation of a single innovator seed may move the dynamical system away from equilibrium and create a global cascade. A linear stability analysis shows that this condition is equivalent to
\begin{equation}
\label{eq:AMEsCond}
(1 - r) \sum_k \frac{k}{z} (k - 1) P(k) f(k, 1) > 1,
\end{equation}
where $f(k, 1) = C(1/k)$ implements the response of a non-immune node of degree $k$ to one adopting neighbour, and $C$ is the cumulative distribution function of $P(\varphi)$ (for non-immune nodes with $c > 0$). When $r = 0$, Eq.~(\ref{eq:AMEsCond}) reduces trivially to the cascade condition of the original WT model in Eq.~(\ref{eq:wattsCond}). This shows 
that the shape of the cascade regime can be obtained either by using generating functions in percolation theory, or by performing a stability analysis of the AMEs.

\section{Validation}
\label{sec:val}

As demonstrated above, our model provides insight on the role of innovators and immune nodes in controlling the speed of the adoption process. However, in empirical datasets information about the fraction of non-adopters is usually not available, which makes it difficult to predict the future dynamics of service adoption. Here we use our modelling framework to perform data-driven simulations with parameters determined from Skype for two reasons: (a) to estimate the fraction $r$ of immune nodes in the real system; and (b) to validate our modelling as compared to real data.

\begin{figure} \centering
\includegraphics[width=0.8\textwidth,angle=0]{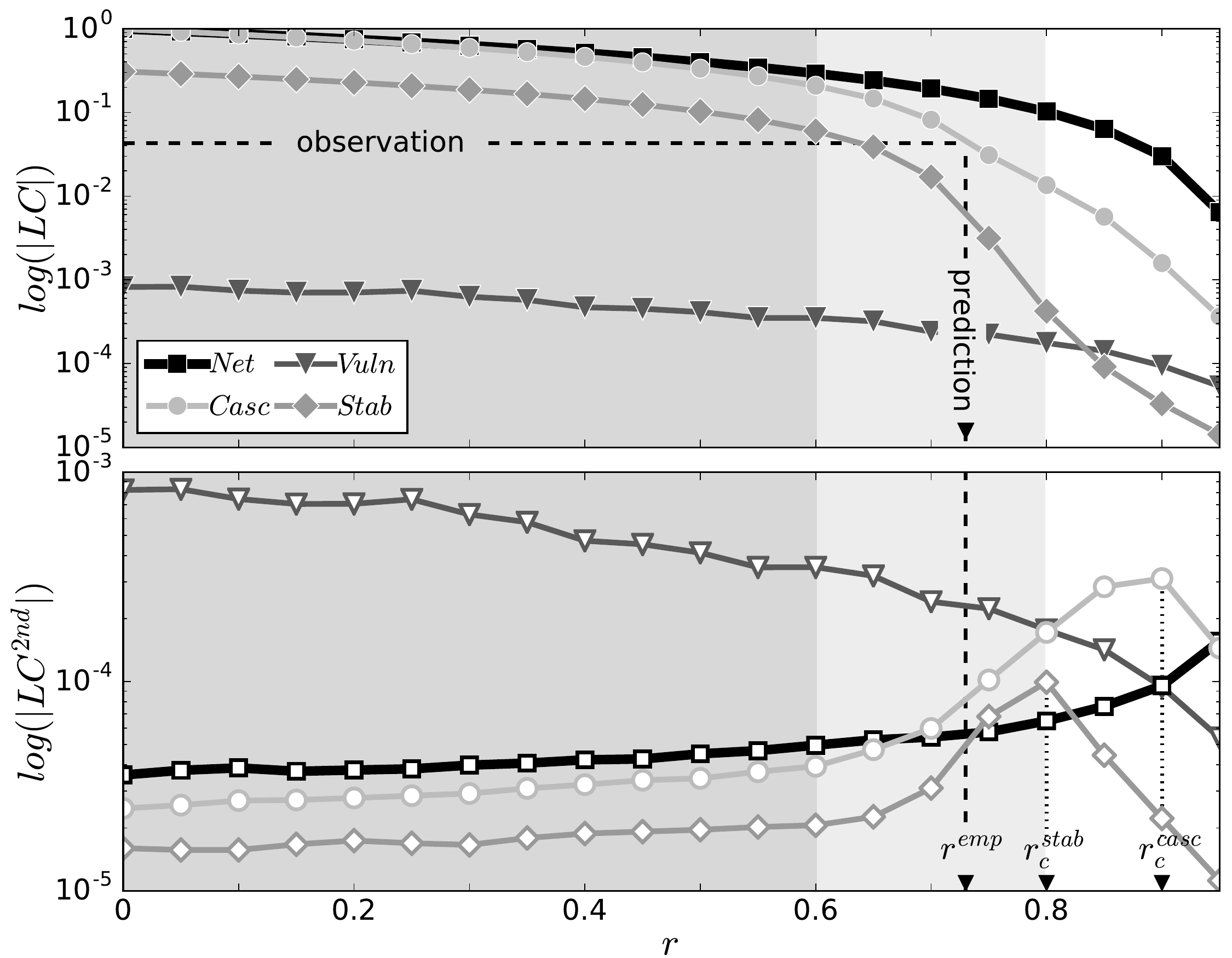}
\caption{\textbf{Empirical cluster statistics and simulation results.} Average size of the largest ($LC$, upper panel) and the 2nd largest ($LC^{2nd}$, lower panel) components of the model network (`Net', squares), adoption network (`Casc', circles), stable network (`Stab', diamonds), and induced vulnerable trees (`Vuln', triangles) as a function of the fraction $r$ of immune nodes. Dashed lines show the observed relative size of the real $LC$ of the adopter network in $2011$ (Fig.\ref{fig:2}c) and the predicted $r^{emp}$ value. Dotted lines on the lower panel indicate the critical percolation points for the full ($r_c^{casc}$) and stable ($r_c^{stab}$) adoption networks.
\label{fig:6}}
\end{figure}

\begin{figure} \centering
\includegraphics[width=\textwidth,angle=0]{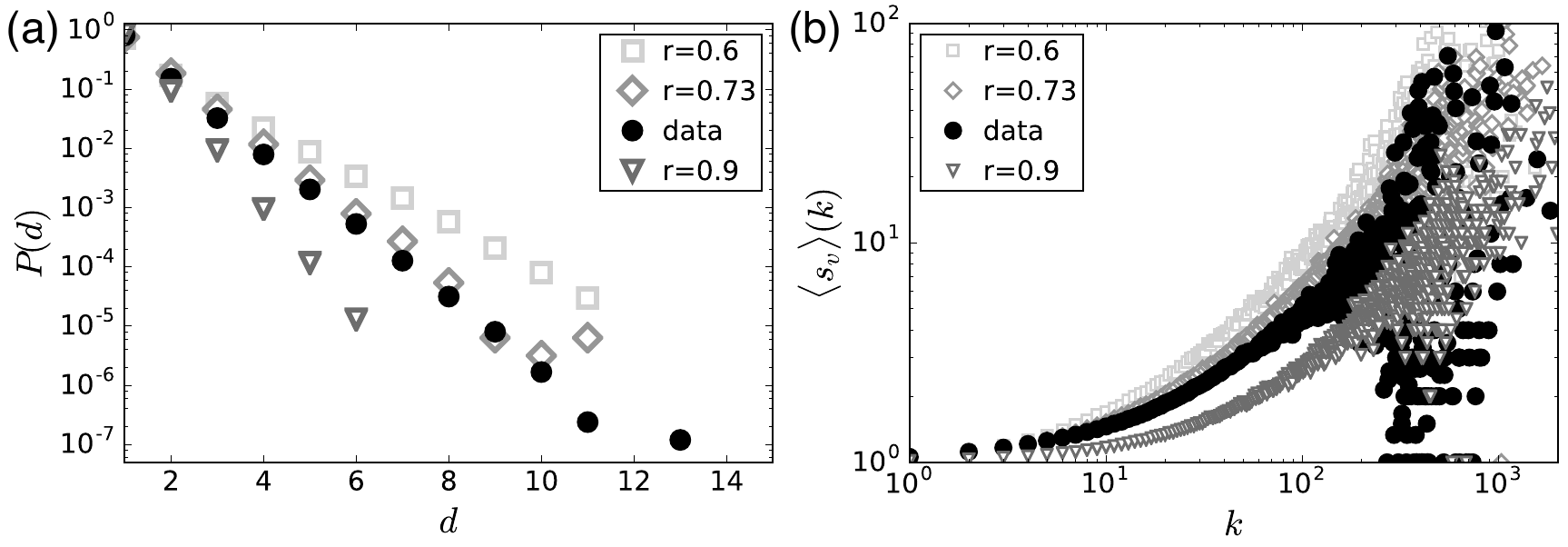}
\caption{\textbf{Additional empirical cluster statistics and simulation results.} \textbf{(a)} Distribution $P(d)$ of depths of induced vulnerable trees in both data and model for several $r$ values, showing a good fit with the data for $r=0.73$. The difference in the tail is due to finite-size effects. \textbf{(b)} Correlation $\langle s_v \rangle (k)$ between innovator degree and average size of vulnerable trees in both data and model with the same $r$ values as in (a). Model calculations correspond to networks of size $N=10^6$ and are averaged over $10^2$ realisations.
\label{fig:7}}
\end{figure}

To set up our data-driven simulations we use the Skype data to directly determine all model parameters, apart from the fraction $r$ of immune nodes. As we already discussed, the best approximation of the degree distribution of the real network is a log-normal function (Eq.~\ref{eq:Pk}) with parameters $\mu_D=1.2$, $\sigma_D=1.39$, minimum degree $\kmin = 1$ and average degree $z = 8.56$. To account for finite-size effects in the model results for low $N$, we decrease $\mu_D$ slightly to obtain the same value of $z$ as in the real network. We also observe in Fig.~\ref{fig:1}b that the threshold distribution of each degree group collapses into a master curve after normalisation by using the scaling relation $P(\Phi_k,k)=k P (\Phi_k/k)$. This master curve can be well-approximated by the log-normal distribution shown in Eq.~\ref{eq:Pphi}, with parameters $\mu_T=-2$ and $\sigma_T=1$ as determined by the empirical average threshold $w = 0.19$ and standard deviation $0.233$. We estimate a rate of innovators $p_n = 0.00019$ by fitting a constant function to $R_i(t)$ for $t > 2\tau$ (Fig.~\ref{fig:2}a). The fit to $\pn$ also matches the time-scale of a Monte Carlo iteration in the model to 1 month. To model the observed dynamics and explore the effect of immune nodes, we use a configuration-model network~\cite{Newman2010Networks} with log-normal degree and threshold distributions and $p_n$ as the constant rate of innovators, all determined from the empirical data. Model results in Fig.~\ref{fig:6} (and Fig.~\ref{fig:7}) are averaged over $100$ networks of size $N=10^5$ ($10^6$) after $T=89$ iterations, matching the length of the observation period in Skype.

As a function of $r$, the underlying and adoption networks pass through three percolation-type phase transitions. First, the appearance of immune nodes (for increasing $r$) can be considered as a removal process of nodes available for adoption from the underlying network structure. After the appearance of a critical fraction of immune nodes, $r_c^{net}$, the effective network structure available for adoption will be fragmented and will consist of small components only, limiting the size of the largest adoption cluster possible. Second, $r$ also controls the size of the emergent adoption cascades evolving on top of the network structure. While for small $r$ the adoption network is connected into a large component, for larger $r$ cascades cannot evolve since there are not enough nodes to fulfill the threshold condition of susceptible stable nodes, even if the underlying network is still connected. The transition point between these two phases of the adoption network is located at $r_c^{casc}\leq r_c^{net}$, limited from above by the critical point $r_c^{net}$. Finally, we observe from the empirical data and model results that the adoption network is held together by a large connected component of stable nodes. Consequently, for increasing $r$ the stable adoption network goes through a percolation transition as well, with a critical point $r_c^{stab}\leq r_c^{casc}\leq r_c^{net}$.

To characterise these percolation phase transitions we compute the average size of the largest ($LC$) and second largest ($LC^{2nd}$) connected components (Fig.~\ref{fig:6}). We measure these quantities for the underlying network, and for the stable, vulnerable and global adoption networks, as a function of the fraction of immune nodes $r$. After $T=89$ iterations (matching the length of the real observation period), we identify three regimes of the dynamics: if $0<r<0.6$ (dark-shaded area) the spreading process is very rapid and evolves as a global cascade, which reaches most of the nodes of the shrinking susceptible network in a few iteration steps. About $10\%$ of adopters are connected in a percolating stable cluster, while vulnerable components remain very small in accordance with empirical observations. In the crossover regime $0.6<r<0.8$ (light-shaded area), the adoption process slows down considerably (Fig.~\ref{fig:6}, upper panel), as stable adoptions become less likely due to the quenching effect of immune nodes. The adoption process becomes the slowest at $r_c^{stab}=0.8$ when the percolating stable cluster falls apart, as demonstrated by a peak in the corresponding $LC^{2nd}$ curve in Fig.~\ref{fig:6} (diamonds in lower panel). Finally, around $r_c^{casc}=0.9$ the adoption network becomes fragmented and no global cascade takes place. Since the underlying network has a broad degree distribution, it is robust against random node removal processes¬\cite{Newman2010Networks}. That is why its critical percolation point $r_c^{net}$ appears after $95\%$ or more nodes are immune. Note that similar calculations for another service have been presented before~\cite{Karsai2016Local} with qualitatively the same results, but with the crossover regime shifted towards larger $r$ due to different parameter values of the model process.

We can use these calculations to estimate the only unknown parameter, namely the fraction $r$ of immune nodes in Skype, by matching the relative size of the largest component ($LC_{Net}$) between real and model adoption networks at time $T$. Empirically, this value is the relative size $s_a^{LC}\simeq 0.043$ corresponding to the last point on the right-hand side of the distribution for $2011$ in Fig.~\ref{fig:2}c (main panel). Matching this relative size with the simulation results (see the observation line in Fig.~\ref{fig:6} upper panel), we find that it corresponds to $r^{emp} = 0.73$ (prediction line in Fig.~\ref{fig:6}), suggesting that the real adoption process lies in the crossover regime. In other words, large adoption cascades could potentially evolve in Skype but with reduced speed, as $73\%$ of users might not be interested in adopting a service within the network.

To test the validity of the predicted $r^{emp}$ value we perform three different calculations. First we measure the maximum relative growth rate of cumulative adoptions and find a good match between model and data (see Skype s3 and Model Skype s3 in Fig.~\ref{fig:0}). In other words, the model correctly estimates the speed of the adoption process. Second, we measure the distribution $P(d)$ of the depths of induced vulnerable trees (Fig.~\ref{fig:7}a). Vulnerable trees evolve with a shallow structure in the empirical and model processes. After measuring the distribution $P(d)$ for various $r$ values below, above and at $r^{emp}$, we find that the distribution corresponding to the predicted $r^{emp}$ value fits the best with the empirical data. Finally, in order to verify earlier theoretical suggestions~\cite{Singh2013Thresholdlimited}, we look at the correlation $\langle s_v \rangle (k)$ between the degree of innovators and the average size of vulnerable trees induced by them (Fig.~\ref{fig:7}b). Similar to the distribution $P(d)$, we perform this measurement on the real data and in the model for $r=0.6$ and $0.9$, as well as for the predicted value $r_{emp}=0.73$. We find a strong positive correlation in the data, explained partially by degree heterogeneities in the underlying social network, but surprisingly well emulated by the model as well. More importantly, although this quantity appears to scale with $r$, the estimated $r$ value fits the empirical data remarkably well, thus validating our estimation method for $r$ based on a matching of relative component sizes.

\section{Conclusion and future directions}
\label{sec:concl}

The analysis and modelling of the diffusion of services and innovations is a long-standing scientific challenge, with recent developments built on large digital datasets registering adoption processes in a society with a large population. Due to these advancements we are currently at the position to simultaneously observe various types of adoption processes and the underlying social structure. Individual-level observations of social and adoption behaviour are crucial in identifying the mechanisms that fuel collective patterns of rapid or slow adoption cascades. In this chapter, using one of the first datasets of this kind, we observe the worldwide spread of an online service in the techno-social communication network of Skype. First we provide novel empirical evidence about heterogeneous adoption thresholds and non-linear dynamics of the adoption process. We have also identified two additional components necessary to introduce into the modelling of product adoption, namely (a) a constant flow of innovators, which may induce rapid adoption cascades even if the system is initially out of the cascading regime, and (b) a fraction of immune nodes that forces the system into a quenched state where adoption slows down. These features are responsible for a critical structure of empirical adoption components that radically differs from previous theoretical expectations. We incorporate these mechanisms into a threshold model that, despite containing several simplifying assumptions, successfully recovers and predicts real-world adoption scenarios such as the spreading of Skype services.

Our aim in this chapter has been to provide empirical observations as well as methods and tools to model the dynamics of social contagion phenomena, with the hope that it will foster thoughts for future research. One possible direction is the observation of the reported structure and evolution of the global adoption cluster in other systems similar to the ones studied in \cite{BorgeHolthoefer2011Structural,Dow2013Anatomy,Gruhl2004Information,Goel2012Structure,BorgeHolthoefer2013Cascading,Bakshy11}. Other promising directions are the consideration of structural homophilic or assortative correlations, the evolving nature of the underpinning social network with timely created and dissolved social ties (as studied in~\cite{Karsai2014Complex}), and the effects of interpersonal influence or leader-follower mechanisms on the social contagion process. We hope that our results provide a direction for data-driven modelling of these phenomena, and serve as a scholarly example in future studies of the dynamics of service adoption processes.

\begin{acknowledgement}
The results presented in this chapter are adapted from~\cite{Ruan2015,Karsai2016Local} and were obtained in collaboration with Riivo Kikas. The authors gratefully acknowledge the support of M. Dumas, A. Saabas, and A. Dumitras from STACC and Microsoft/Skype Labs. GI acknowledges a Visiting Fellowship from the Aalto Science Institute. JK and ZR were supported by FP7 317532 Multiplex and JK by H2020 FETPROACT-GSS CIMPLEX 641191. KK is supported by the Academy of Finland's project COSDYN project, No. 276439 and EU HORIZON 2020 FET Open RIA IBSEN project No. 662725.
\end{acknowledgement}

\end{document}